# Giant and anisotropic many-body spin-orbit tunability in a strongly correlated kagome magnet

**Authors:** Jia-Xin Yin[1]*, Songtian S. Zhang[1]*, Hang Li[2]*, Kun Jiang[3], Guoqing Chang[1], Bingjing Zhang[4], Biao Lian[5], Cheng Xiang[6,7], Ilya Belopolski[1], Hao Zheng[1], Tyler A. Cochran[1], Su-Yang Xu[1], Guang Bian[1], Kai Liu[4], Tay-Rong Chang[8], Hsin Lin[9], Zhong-Yi Lu[4], Ziqiang Wang[3], Shuang Jia[6,7], Wenhong Wang[2], M. Zahid Hasan[1,10]†

**Affiliations:**

[1]Laboratory for Topological Quantum Matter and Advanced Spectroscopy (B7), Department of Physics, Princeton University, Princeton, New Jersey 08544, USA.

[2]Beijing National Laboratory for Condensed Matter Physics, Institute of Physics, Chinese Academy of Sciences, Beijing 100190, China.

[3]Department of Physics, Boston College, Chestnut Hill, Massachusetts 02467, USA.

[4]Department of Physics and Beijing Key Laboratory of Opto-electronic Functional Materials & Micro-nano Devices, Renmin University of China, Beijing 100872, China.

[5]Princeton Center for Theoretical Science, Princeton University, Princeton, New Jersey 08544, USA.

[6]International Center for Quantum Materials and School of Physics, Peking University, Beijing 100871, China.

[7]CAS Center for Excellence in Topological Quantum Computation, University of Chinese Academy of Science, Beijing 100190, China.

[8]Department of Physics, National Cheng Kung University, Tainan 701, Taiwan.

[9]Institute of Physics, Academia Sinica, Taipei 11529, Taiwan.

[10]Lawrence Berkeley National Laboratory, Berkeley, California 94720, USA.

†Corresponding author, E-mail: mzhasan@princeton.edu

*These authors contributed equally to this work.

**Owing to the unusual geometry of kagome lattices—lattices made of corner-sharing triangles—their electrons are useful for studying the physics of frustrated, correlated and topological quantum electronic states[1-9]. In the presence of strong spin–orbit coupling, the magnetic and electronic structures of kagome lattices are further entangled, which can lead to hitherto unknown spin–orbit phenomena. Here we use a combination of vector-magnetic-field capability and scanning tunnelling microscopy to elucidate the spin–orbit nature of the kagome ferromagnet $Fe_3Sn_2$ and explore the associated exotic correlated phenomena. We discover that a many-body electronic state from the kagome lattice couples strongly to the vector field with three-dimensional anisotropy, exhibiting a magnetization-driven giant nematic (two-fold-symmetric) energy shift. Probing the fermionic quasi-particle interference reveals consistent spontaneous nematicity—a clear indication of electron correlation—and vector magnetization is**

**capable of altering this state, thus controlling the many-body electronic symmetry. These spin-driven giant electronic responses go well beyond Zeeman physics and point to the realization of an underlying correlated magnetic topological phase. The tunability of this kagome magnet reveals a strong interplay between an externally applied field, electronic excitations and nematicity, providing new ways of controlling spin–orbit properties and exploring emergent phenomena in topological or quantum materials[10-12].**

Understanding and manipulating correlated quantum materials are prerequisites for finding and exploring their potential for applications[10–12] and quantum materials that exhibit a giant response in the presence of an external field are particularly promising[10–12]. Kagome antiferromagnets are central in the search for exotic quantum states because both the spin and the charge are frustrated geometrically, enabling the formation of spin-liquid phases and topological electronic structures[1–9]. However, the realization of such states in real materials has been limited. Kagome ferromagnets are also of great interest because their unusual physics can be probed in transport, such as in transition-metal stannides[13–18]. Transport measurements in this family have demonstrated large anomalous Hall effects that can arise from nontrivial electronic topology with non-vanishing Berry curvatures[13–18]. The Berry phase of the antiferromagnet $Mn_3Sn$ is associated with a non-collinear spin texture and the existence of topological fermions in its band structure[15–17]. For the soft ferromagnet $Fe_3Sn_2$, it is speculated that the Berry phase is associated with a massive Dirac band near the corner of the Brillouin zone, which hosts a two-dimensional gap[18]. Accordingly, this family serves as a fertile platform for exploring the interplay between magnetism and quantum electronic structure in kagome lattices. Here we study the atomically resolved electronic structure of $Fe_3Sn_2$ at 4.2 K (Curie temperature, $T_{Curie}$ = 670 K) by using a combination of time-reversal-breaking vector magnetic-field microscopy and high-resolution scanning tunnelling microscopy/spectroscopy (STM/S). Although many previous works focused on the unusual transport properties of $Fe_3Sn_2$, we observe an unexpected giant anisotropic vector-field response of the electronic states of the kagome lattice, which opens up the opportunity to demonstrate controlled quantum-level manipulation of an exotic topological phase. The methodology described here offers a new way of discovering magnetic topological phases in a strongly correlated setting, which can be used for the discovery of other correlated topological materials.

$Fe_3Sn_2$ has a layered crystal structure with space group $R\bar{3}m$ and hexagonal lattice constants a = 5.3 Å and c = 19.8 Å. It consists of a honeycomb Sn layer sandwiched between kagome FeSn bilayers (Fig. 1a). Due to their weak bonding, the sample tends to cleave with either a FeSn or Sn terminated surface. These two surfaces are experimentally identified via comparisons of their respective step edge heights in the crystal structure (Fig. 1b). Mapping the differential conductance of these two surfaces also reveals differences in the electronic structure (Fig. 1c). A detailed inspection of Fig. 1d confirms the honeycomb lattice structure of the Sn surface, while the FeSn surface exhibits smaller corrugation, hindering direct atomic identification. By analysing the line-cuts taken across the step edge of the Sn surface (Fig. 1e), we assign the positions of the Sn atom, corresponding to the centre of the Sn honeycomb unit, and the Fe atoms in the FeSn surface (Fig. 1d).

Having identified the two surfaces, we study the quasiparticle excitations under the perturbation of an external magnetic field. At zero field, the low energy differential conductance spectra on both surfaces show strong states around the Fermi energy with the spectrum on the FeSn surface exhibiting an additional state at -8mV. Increasing the c-axis field causes a pronounced shift of the side peak towards negative energies, while there is no discernible shift of the states near the Fermi energy. From the magnetic response and surface dependence, it is likely that the side peak arises from the magnetic Fe orbital in the kagome lattice. We find that its magnetic field response extends beyond Zeeman effect in several aspects. The shift of the side peak saturates around 1T, with a total energy-shift of 12meV (Fig. 2a). We observe an identical shift when the 1T field is reversed. More importantly, the saturation behaviour agrees well with the magnetization curve (Fig. 2c), denoting a magnetization driven shift. If the energy-shift is attributed to the Zeeman effect, it would amount to an anomalously large value for the *g* factor, not known in the previous literature (Fig. 2b).

To explore its magnetization response in three dimensions, we rotate the external field in the ab plane. We find that this state from the FeSn surface also saturates before 1T when the field is applied in-plane, and that the saturated shift with 1T field evolves with the azimuth angle θ (Fig. 2d). In contrast to the six-fold crystal symmetry, the evolution has a two-fold symmetry, which can be described by the function: $3.2-3.2\cos 2\theta$ meV (Fig. 2e). Notably, there is no shift when the field is applied along the a-axis (θ = 0). As the net magnetization is known to lie in plane at low temperatures in zero field[19,20], such nodal behaviour can be understood by considering the spontaneous magnetization to be along the a-axis, which already saturates the energy-shift. The anisotropic evolution we observe in our STM data also qualitatively agrees with the bulk transport anisotropy in response to a vector field (Extended Data Fig. 2), consistently demonstrating the existence of electronic nematicity in $Fe_3Sn_2$, previously observed in correlated materials[21-24].

To further study the symmetry of the electronic state realised in this material, we map the differential conductance of the FeSn surface over a large area under various vector field conditions. Taking their Fourier transforms, we obtain the quasiparticle interference (QPI) data for the electron scattering involving the band structure. The QPI data taken at the energy of this side peak state are shown in Fig. 3. The zero-field QPI data in Fig. 3a exhibits ring-like signals at the larger wave vectors (q2) and a two-fold pattern around the zone centre (q1). This spontaneously broken symmetry state also aligns with the sample a-axis, in agreement with the aforementioned nematicity, consistent with our transport data. Remarkably, we find that the c-axis magnetization removes the electronic nematicity and restores the rotational symmetry (Fig. 3b). While the a-axis magnetization retains the same nematic pattern at q1 (Fig. 3c), this pattern is systematically rotated by the rotation of in-plane magnetization. When the external field is withdrawn, the nematicity recovers to that in Fig. 3a regardless of magnetization history that we checked. This suggests that there exists an intrinsic nematic order pinning the spontaneous magnetization to be along the a-axis. In contrast, the QPI around q2 remains approximately isotropic regardless of field, indicating that the shifting of the electronic state is most likely associated with the states spanning the momentum transfer q1.

Such an association is further supported by the field dependent QPI dispersion plotted in Fig. 3d and e. The QPI dispersions in Fig. 3d shows a clear hole-like band (q2) with no discernible field dependence, corresponding to the ring like signal seen in the QPI images in Fig. 3a-c. On the other hand, at small q, an examination of the dispersion reveals a maximum in the QPI intensity at approximately the energy of the side peak for both the zero-field (E = -10 ± 2mV) and B=1T (E = -20 ± 2mV) case. This QPI intensity extends to larger q with increasing energy, suggesting an electron-like band with a band bottom at this low q maximum energy peak (q1). Moreover, when the energy window of the dispersion is extended to even lower energies (Fig. 3e), a second hole-like band appears at B = 1T, with these two upper and lower branches forming a non-monotonically dispersing feature which resembles the hourglass shape. The non-monotonic shape of the observed signal, despite their broadness, indicates that the scattering is sensitive to the details of the underlying dispersion in the band structure. It is consistent with a massive Dirac-like dispersion, under the assumption that the scattering is intra-band in nature and corresponds to the Dirac feature expected around this corresponding energy in the photoemission measurement[18]. These two dispersive branches are farther apart from the centre of the hourglass feature as one tunes the magnetic field to zero. Recent identification of band structure resembling massive Dirac fermions in this material based on photoemission suggests a gap size of about 30 mV, consistent with our data at B = 0T in Fig. 3e, despite what appears to be a shift in the chemical potential, possibly due to surface termination effects or doping differences in the samples. It is well known that photoemission measurements lack the ability to probe the field dependence of this gap (mass of the Dirac band) or determine whether the Dirac fermions originate from the FeSn kagome lattice or Sn honeycomb lattice which is critically important to correctly model or theoretically understand the novel state realized in this material.

A summary of our experimental findings is shown in Fig. 4. Our results establish a vector magnetization based energy-shift of the quantum electronic states with intriguing symmetry-breaking correspondence (Fig. 4a and b). These states form an electron band crossing the Fermi level (Fig. 4c). Without the external field, the spontaneous magnetization is along the a-axis; the band bottom (identified with the side peak in the tunnelling conductance in Fig. 2a) exhibits a QPI with a two-fold symmetry. While the symmetry of the QPI rotates with the magnetization indicating strong spin-orbit coupling (SOC) that intertwines the orbital space with magnetic space, it is unexpected that the energy of the band bottom modulates significantly with the angle of rotation. The observed in-plane magnetization induced energy-shift also has a two-fold symmetry with its nodal line along the a-axis (Fig. 4a), indicative of an intrinsic nematic order that pins the spontaneous magnetization direction and leads to this energy difference. Rotating the magnetization to the c-axis causes the largest energy-shift (Fig. 4a). These giant electronic responses driven by the magnetization direction goes well beyond Zeeman physics, and points to a spin-orbit entangled, correlated magnetic topological phase, which we discuss below.

In fact, previous STM work on other systems has shown that due to the presence of SOC, the electronic structure of magnetic thin films with domain walls[25] and skyrmions[26] can have dependence on the spin orientation. As electronic structures on the kagome lattice possess linear band crossing Dirac points at the Brillouin zone corners, it is natural for us to consider a picture of Dirac fermions in the presence of SOC in the study of kagome lattice (quantum) anomalous Hall materials[8,13,14]. The observed energy-shift should thus result from the interplay of the Dirac mass/gap (Fig.4c) and magnetism. The large ferromagnetic moment splits the Dirac crossing into two sets well separated in energy with spins polarized in the direction of magnetization. In Ref. 18, the magnetization is assumed to be along the c-axis, and a Kane-Mele type[27] SOC that preserves the spin $S_Z$ component is considered that produces a Dirac mass/gap. However, the spontaneous magnetization in $Fe_3Sn_2$ lies in the plane at low temperatures such that the $S_Z$ SOC cannot gap out the Dirac crossing in the kagome lattice[28]. This contradicts our observation of the largest mass gap (smallest energy-shift) for a-axis magnetization sketched in Figs. 4a and 4c. Thus, the physics governing the interplay between SOC and magnetism here lies beyond the Kane-Mele scenario. One possibility is that all SOC interactions respecting the full crystal symmetry need to be constructed with both $S_Z$ and in-plane $S_{X,Y}$ conserving terms, where our results indicate that the latter should have a larger effect. Alternatively, an additional source for the Dirac mass/gap may exist and interfere with the one due to a dominant Kane-Mele SOC. Since $Fe_3Sn_2$ displays a large anomalous Hall effect with skyrmion excitations[20], it is possible that the latter has contribution from the spin Berry phase[3] associated with chiral spin textures[19]. The spin chirality produces a gauge flux, according to the theory[3], opens a Dirac gap independent of the magnetization direction. As the magnetization is rotated to the c-axis, the orbital flux induced by the Kane-Mele SOC can be out of phase and compete with the gauge flux, leading to the reduction of the Dirac mass/gap (Extended Data Fig. 11), consistent with our interpretation of the data.

Our experiment furthermore reveals an intriguing nematic order in this kagome magnet. In addition to the magnetization controlled charge nematicity due to SOC effects, there exists an intrinsic nematic order originating from the charge channel, as evidenced by the anisotropic energy-shift and transport response to the vector magnetization, and the pinning of the spontaneous magnetization direction irrespective of vector magnetization history. Interestingly, the well-known intra-unit-cell (q=0) charge ordered state on the kagome lattice driven by intersite Coulomb interactions is a nematic state as demonstrated in theory[29,30].

In summary, our experiment uncovers a vector field based energy-shift to broken-symmetry correspondence in $Fe_3Sn_2$, which demonstrates unusually large and anisotropic magnetic tunability in a spin-orbit kagome magnet and points to an underlying correlated magnetic topological ground state. The novelty in the current work regards the spin-orbit tunability and the gigantic response of the kagome material, and such an effect is not implied by, nor can it be derived from, the previously known transport or photoemission results. The gigantic spin-orbit response we discovered in this strongly correlated material is rather unexpected and not implied by results reported in Ref-13, 14, 18. Our findings collectively show the rich and unconventional physics of kagome magnets, encompassing the entangled magnetic, charge, and orbital degree of freedom, as well as symmetry breaking and topological properties of the electronic states involving low-energy fermions. A complete understanding would require a comprehensive quantum many-body theory of electrons on the kagome lattice in the presence of strong spin-orbit coupling. Our space-momentum exploration of the electronic excitations by way of controlled vector field manipulation presents itself as a new and powerful tool for probing the physics of topological matter beyond weakly interacting $Z_2$ topological insulators[27].

**References:**

1. Mekata, Mamoru. Kagome: The story of the basketweave lattice. *Physics Today* **56**, 12 (2003).

2. Zhou, Y., Kanoda, K. & Ng, T. -K. Quantum spin liquid states. *Rev. Mod. Phys.* **89**, 025003 (2017).

3. Ohgushi, K., Murakami, S. & Nagaosa, N. Spin anisotropy and quantum Hall effect in the kagomé lattice: Chiral spin state based on a ferromagnet. *Phys. Rev. B* **62**, R6065(R) (2000).

4. Yan, S., Huse, D. A. & White. S. R. Spin-Liquid Ground State of the S=1/2 Kagome Heisenberg Antiferromagnet. Science 332, 1173-1176 (2011).

5. Han, T. -H. *et al.* Fractionalized excitations in the spin-liquid state of a kagome-lattice antiferromagnet. *Nature* **492**, 406-410 (2012).

6. Mazin, I. I. *et al.* Theoretical prediction of a strongly correlated Dirac metal. *Nat. Comm.* **5**, 4261 (2014).

7. Chisnell, R. *et al.* Topological magnon bands in a kagome lattice ferromagnet. *Phys. Rev. Lett.* **115**, 147201 (2015).

8. Xu, G., Lian, B. & Zhang, S.-C. Intrinsic quantum anomalous Hall effect in the kagome lattice $Cs_2LiMn_3F_{12}$. *Phys. Rev. Lett.* **115**, 186802 (2015).

9. Zhu, W., Gong, S.-S., Zeng, T.-S., Fu, L. & Sheng, D. S. Interaction-driven spontaneous quantum Hall effect on a kagome lattice. *Phys. Rev. Lett.* **117**, 096402 (2016).

10. Soumyanarayanan, A., Reyren, N., Fert, A. & Panagopoulos, C. Emergent Phenomena Induced by Spin-Orbit Coupling at Surfaces and Interfaces. *Nature* **539**, 509-517 (2016).

11. Keimer, B. & Moore, J. E. The physics of quantum materials. *Nature Physics* **13**, 1045-1055 (2017).

12. Tokura, Y., Kawasaki, M. & Nagaosa, N. Emergent functions of quantum materials. *Nature Physics* **13**, 1056-1068 (2017).

13. Kida, T. *et al.* The giant anomalous Hall effect in the ferromagnet $Fe_3Sn_2$ - a frustrated kagome metal. *J. Phys.: Condens. Matter* **23**, 112205 (2011).

14. Wang, Q., Sun, S., Zhang, X., Pang, F. & Lei, H. Anomalous Hall effect in a ferromagnetic $Fe_3Sn_2$ single crystal with a geometrically frustrated Fe bilayer kagome lattice. *Phys. Rev. B* **94**, 075135 (2016).

15. Nakatsuji, S., Kiyohara, N. & Higo, T. Large anomalous Hall effect in a non-collinear antiferromagnet at room temperature. *Nature* **527**, 212-215 (2015).

16. Nayak, A. K. *et al.* Large anomalous Hall effect driven by a nonvanishing Berry curvature in the noncolinear antiferromagnet $Mn_3Ge$. *Science Advances* **2**, e1501870 (2016).

17. Kuroda, K. *et al.* Evidence for magnetic Weyl fermions in a correlated metal. *Nat. Mater.* **16**, 1090-1095 (2017).

18. Ye, Linda *et al.* Massive Dirac fermions in a ferromagnetic kagome metal. *Nature* **555**, 638-642

(2018).

19. Fenner, L. A., Dee, A. A. & Wills, A. S. Non-collinearity and spin frustration in the itinerant kagome ferromagnet $Fe_3Sn_2$. *J. Phys.: Condens. Matter* **21**, 452202 (2009).

20. Hou, Z. *et al.* Observation of various and spontaneous magnetic Skyrmionic bubbles at room temperature in a frustrated kagome magnet with uniaxial magnetic anisotropy. *Adv. Mater.* **29**, 1701144 (2017).

21. Fradkin, E., Kivelson, S. A., Lawler, M. J., Eisenstein, J. P. & Mackenzie, A. P. Nematic Fermi Fluids in Condensed Matter Physics. *Annual Review of Condensed Matter Physics* **1**, 153-178 (2010).

22. Borzi, R. A. *et al.* Formation of a nematic fluid at high fields in $Sr_3Ru_2O_7$. *Science* **315**, 214-217 (2007).

23. Chuang, T. M. *et al.* Nematic Electronic Structure in the "Parent" State of the Iron Based Superconductor $Ca(Fe_{1-x}Co_x)_2As_2$. *Science* **327**, 181-184 (2010).

24. Fujita, K. *et al.* Simultaneous transitions in cuprate momentum-space topology and electronic symmetry breaking. *Science* **344**, 612-616 (2014).

25. Bode, M. *et al.* Magnetization-direction-dependent local electronic structure probed by scanning tunneling spectroscopy. *Phys. Rev. Lett.* **89**, 237205 (2002).

26. Hanneken, Christian *et al.* Electrical detection of magnetic skyrmions by tunnelling non-collinear magnetoresistance. *Nature Nanotechnology* **10**, 1039-1042 (2015).

27. Kane, C. L. & Mele, E. J. Quantum spin Hall effect in graphene. *Phys. Rev. Lett.* **95**, 226801 (2005).

28. Hasan, M. Z. & Kane, C. L. Colloquium: Topological insulators. *Rev. Mod. Phys.* **82**, 3045 (2010).

29. Guo, H. M. & Franz, M. Topological insulator on the kagome lattice. *Phys. Rev. B* **80**, 113102 (2009).

30. Nishimoto, S. *et al.* Metal-insulator transition of fermions on a kagome lattice at 1/3 filling. *Phys. Rev. Lett.* **104**, 196401 (2010).

**Acknowledgments:** Experimental and theoretical work at Princeton University was supported by the Gordon and Betty Moore Foundation (GBMF4547/Hasan) and the United States Department of Energy (US DOE) under the Basic Energy Sciences programme (grant number DOE/BES DE-FG-02-05ER46200). Work at the Institute of Physics of the Chinese Academy of Science (IOP CAS) was supported by the National Key R&D Program of China (grant number 2017YFA0206303). Work at Boston College is supported by US DOE grant DE-FG02-99ER45747. We also acknowledge the Natural Science Foundation of China (grant numbers 11790313, 11774422 and 11774424), National Key R&D Program of China (numbers 2016YFA0300403 and 2017YFA0302903), the Key Research Program of the Chinese Academy of Sciences (number XDPB08-1), Princeton Center for Theoretical

Science (PCTS) and Princeton Institute for the Science and Technology of Materials (PRISM)'s Imaging and Analysis Center at Princeton University. T.-R.C. was supported by the Ministry of Science and Technology under a MOST Young Scholar Fellowship (MOST Grant for the Columbus Program number 107-2636-M-006-004-), the National Cheng Kung University, Taiwan, and the National Center for Theoretical Sciences (NCTS), Taiwan. M.Z.H. acknowledges support from Lawrence Berkeley National Laboratory and the Miller Institute of Basic Research in Science at the University of California, Berkeley in the form of a Visiting Miller Professorship. We thank D. Huse and T. Neupert for discussions.

**Author contributions:**

J.-X.Y. and S.S.Z. conducted the STM and STS experiments in consultation with M.Z.H.; H.L.,W.W., C.X. and S.J. synthesized and characterized the sequence of samples; K.J., G.C., B.Z., B.L., K.L., T-R.C., H.L., Z.-Y.L. and Z.W. carried out theoretical analysis in consultation with J.Y. and M.Z.H.; I.B., T.A.C., H.Z., S.-Y.X. and G.B. contributed to sample characterization and instrument calibration; J.-X.Y., S.S.Z. and M.Z.H performed the data analysis and figure development and wrote the paper with contributions from all authors; M.Z.H. supervised the project. All authors discussed the results, interpretation and conclusion.

**Main Figures:**

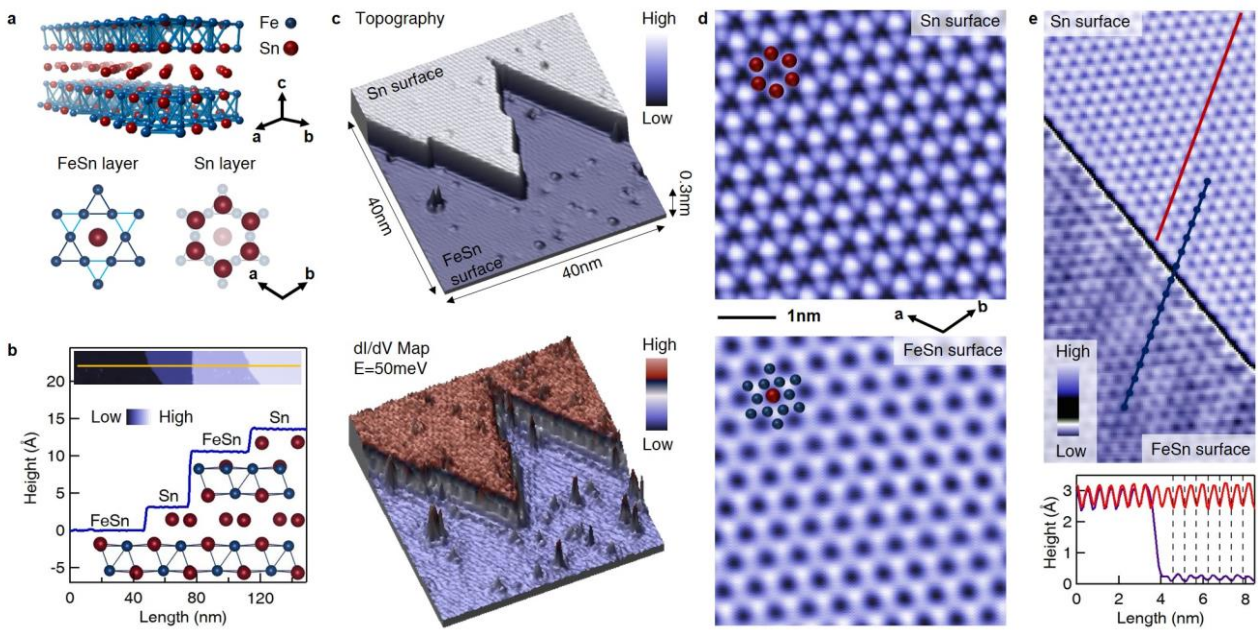

**Figure 1. Surface identification at the atomic scale. a,** Crystal structure of $Fe_3Sn_2$. The lower panels illustrate the kagome lattice of the FeSn layer and the honeycomb lattice of the Sn layer. **b,** Atomic steps created by cryogenic cleaving. From comparison of the step edge height and c-axis crystal structure we can determine the FeSn and Sn surfaces. The inset shows the topographic image with multiple steps (V = 50meV, I = 0.03nA). **c,** Topographic image of a single atomic step with its corresponding differential conductance mapping at 50meV. **d,** Atomically resolved Sn and FeSn surfaces with atoms matched to the crystal structure as marked, respectively (V = 50meV, I = 0.8nA). **e**, Lattice alignment between Sn surface and FeSn surface from a Sn surface step edge. The lower panel compares the two line-cut profiles.

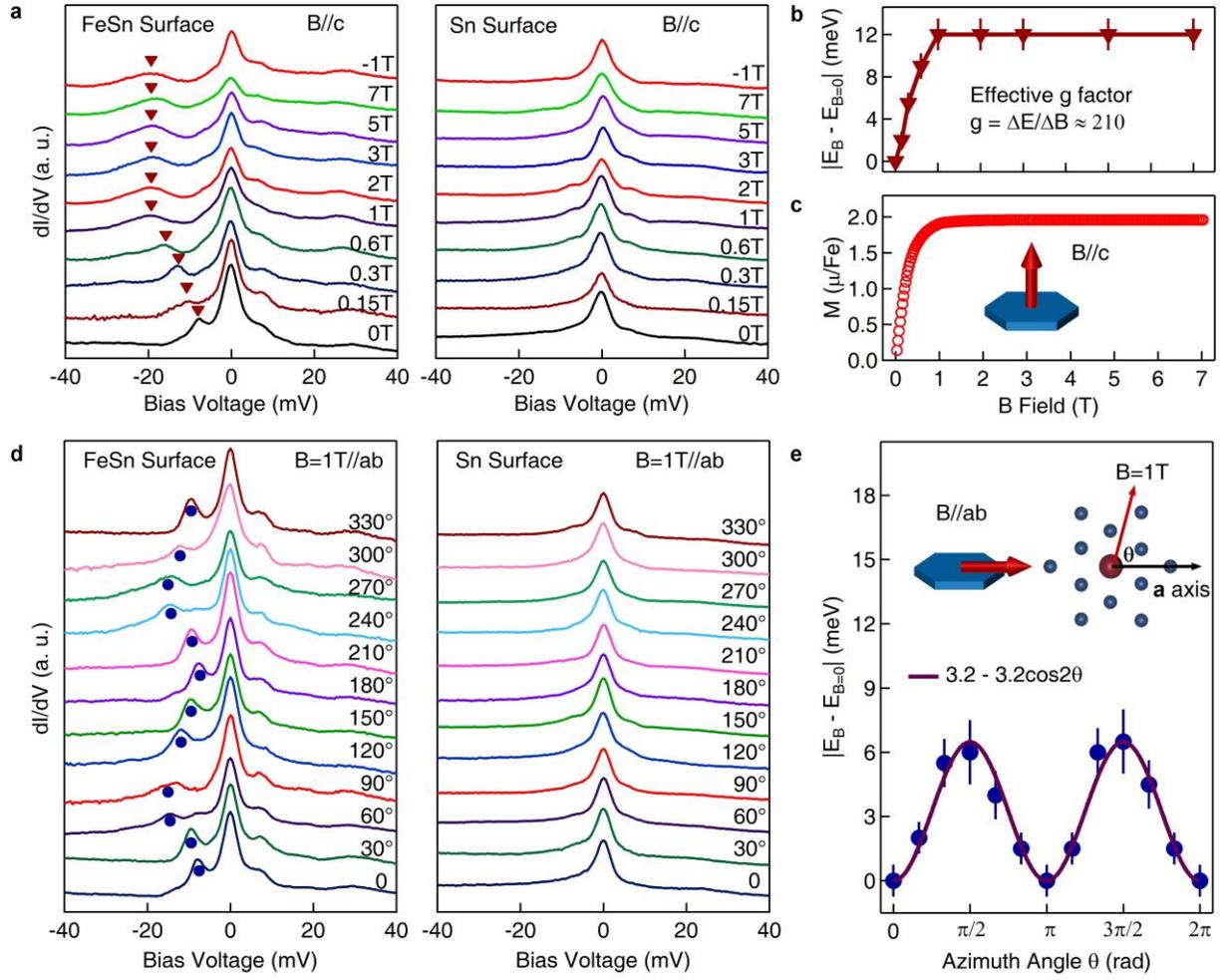

**Figure. 2 Vector magnetization induced giant and nematic energy-shift. a,** c-axis field dependent differential spectra taken on FeSn and Sn surface, respectively. Spectra are offset for clarity. **b,** Energy-shift of the electronic state from FeSn surface as a function of c-axis field. From the initial shift rate below 1T, we can derive the effective g factor around 210. **c**, Bulk c-axis magnetization curve, which correlates strongly with the energy-shift. **d,** In-plane field (B = 1T) angle dependent differential spectra taken on $Fe_3Sn$ and Sn surface, respectively. **e,** Energy-shift as a function of azimuth angle, which can be fitted by a two-fold symmetric function as $3.2-3.2\cos2\theta$. The inset image illustrates the field azimuth angle with respect to the kagome lattice a-axis.

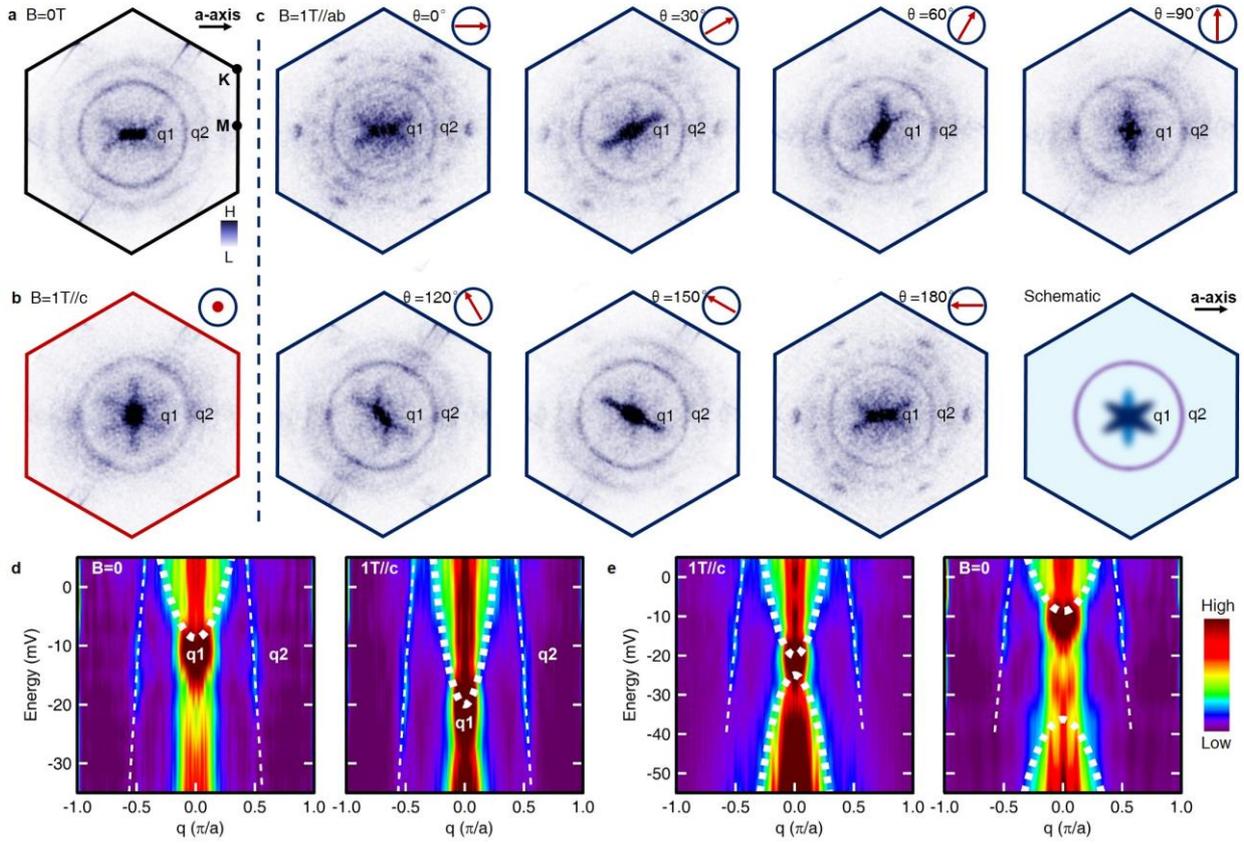

**Figure. 3 Vector magnetization governed electronic symmetry. a,** QPI data taken on the FeSn at the side peak energy with zero field, showing a ring like signal at the wave vector q2 and a two-fold symmetric pattern along the a-axis at the wave vector q1. **b,** QPI data taken with a 1T c-axis field, showing a six-fold symmetric pattern at q1. **c,** QPI data taken under a 1T in-plane vector field showing systematic rotating of the pattern at q1. The bottom right panel illustrates the dominating scattering vectors of the observed QPI signals. The red arrows illustrate the applied field direction, while the black arrow marks the spontaneous magnetization direction. **d,** Field dependent QPI dispersion along the a-axis (Γ-M direction). **e,** QPI dispersion along the a-axis (Γ-M direction) with a wider energy range showing signatures of both the upper and lower branches of the massive Dirac band. The thin dashed lines illustrate the field independent hole-like band (q2) dispersion. The thick dashed lines illustrate a possible field dependent massive Dirac dispersion, which heuristically fits to the intensity fade away of the QPI signal around q=0. We note that all the QPI data are unsymmetrized.

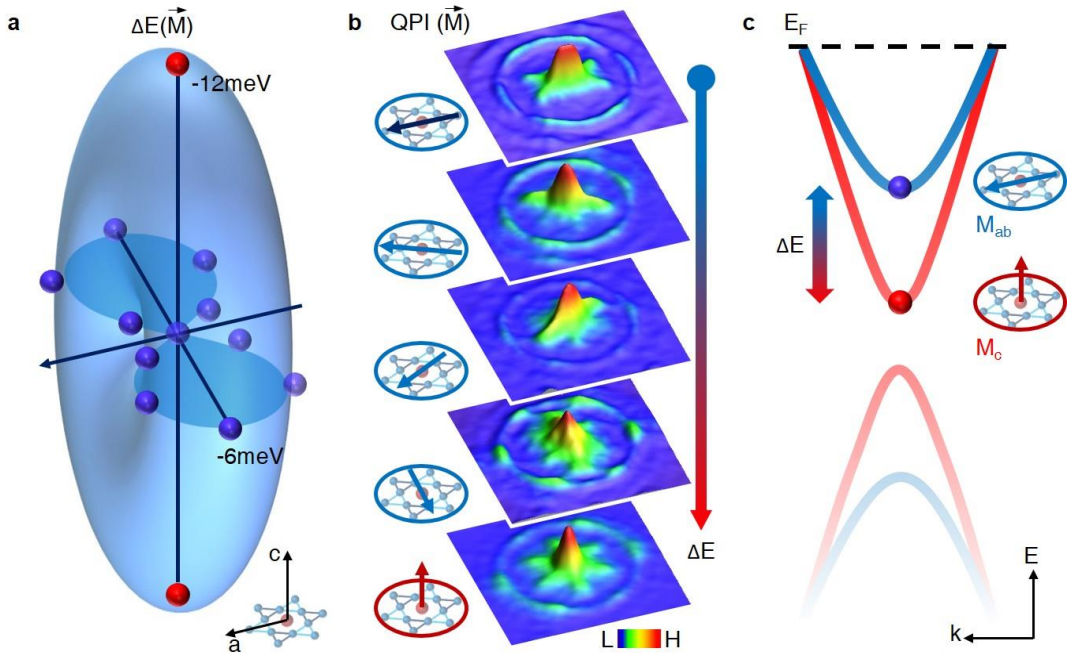

**Figure. 4 Vector magnetization based energy-shift to broken-symmetry correspondence. a,** Saturated energy-shift $\Delta E = E_B - E_{B=0}$ of the electronic state from kagome lattice as a function of magnetization direction vector. The red and blue dots are data from Fig 2a and d respectively. The light blue surface is its 3D illustration, exhibiting a nodal line along the a-axis. **b,** QPI patterns as a function of magnetization direction, indicated by the arrows with respect to the kagome lattice. The upper most QPI shows the spontaneous nematicity along the a-axis. Magnetization along other directions can alter thus control the electronic symmetry. **c,** Schematic of the magnetization controlled Dirac mass/gap, with the band bottom of the upper branch corresponding to the shifting state with broken symmetry. The bands lose coherence away from $E_F$.